\documentclass[letterpaper]{article} 
\usepackage{aaai2026}  
\usepackage{times}  
\usepackage{helvet}  
\usepackage{courier}  
\usepackage[hyphens]{url}  
\usepackage{graphicx} 
\usepackage{natbib}  
\usepackage{caption} 
\usepackage{amsmath}
\usepackage{booktabs}
\usepackage{multirow}
\usepackage{makecell}
\usepackage{subcaption}
\usepackage{color}
\usepackage{amssymb}
\usepackage{algorithm}
\usepackage{algorithmic}

\usepackage{newfloat}
\usepackage{listings}
\DeclareCaptionStyle{ruled}{labelfont=normalfont,labelsep=colon,strut=off} 
\floatstyle{ruled}
\newfloat{listing}{tb}{lst}{}
\floatname{listing}{Listing}
\nocopyright 

\title{\textit{EC$^2$MoE}: Adaptive End-Cloud Pipeline Collaboration Enabling Scalable Mixture-of-Experts Inference}

\author{
   Zheming Yang\textsuperscript{\rm 1}, Yunqing Hu\textsuperscript{\rm 1,3}, Sheng Sun\textsuperscript{\rm 1}, Wen Ji\textsuperscript{\rm 1,2}
   }
\affiliations{
    
    \textsuperscript{\rm 1}Institute of Computing Technology, Chinese Academy of Sciences\\
   \textsuperscript{\rm 2}Institute of Al for Industries \\
  \textsuperscript{\rm 3}University of Chinese Academy of Sciences

}

\usepackage{bibentry}

\begin{document}

\maketitle

\begin{abstract}
The Mixture-of-Experts (MoE) paradigm has emerged as a promising solution to scale up model capacity while maintaining inference efficiency. However, deploying MoE models across heterogeneous end-cloud environments poses new challenges in expert scheduling, communication overhead, and resource heterogeneity. In this paper, we propose \textit{EC$^2$MoE}, an adaptive framework for scalable MoE inference via end-cloud pipeline collaboration. First, we design a hardware-aware lightweight group gate network that enhances expert selection and computational efficiency. By incorporating a hardware-aware local expert selection mechanism, the system adaptively filters candidate experts based on real-time device profiles. A lightweight group gate module then integrates local and global gating outputs to achieve high-quality expert routing with minimal overhead. Second, we develop a pipeline optimization mechanism based on end-cloud collaboration to accelerate MoE inference. This includes an encoder-decoder structure based on low-rank compression, which reduces transmission and computation costs. And a route-aware heuristic pipeline scheduling algorithm that dynamically allocates inference stages across devices according to workload and network topology. Extensive experiments show that \textit{EC$^2$MoE} can increase throughput by 2.2$\times$ to 5.1$\times$ and reduce end-to-end latency by 53\% to 67\% while maintaining high accuracy compared to state-of-the-art methods. It also maintains good scalability under dynamic load and network environments.
\end{abstract}


\section{Introduction}
\label{sec:intro}

In recent years, the demand for large-scale deep learning models has grown dramatically \cite{ge2023openagi111,shen2024efficient222,menghani2023efficient333}, driven by the rapid advancement of AI applications in areas such as natural language understanding, computer vision, and multi-modal reasoning \cite{liang2024survey444}. To meet this demand, researchers have explored various model scaling strategies \cite{hwang2023tutel555,chen2024more666,yin2024enhancing7777}. Among them, the Mixture-of-Experts (MoE) architecture has emerged as a particularly promising solution \cite{zhou2022mixture888}. By selectively activating a sparse subset of expert networks during inference, MoE models enable substantial increases in parameter count—often reaching hundreds of billions—without incurring a proportional increase in computational cost \cite{chen2022towards999,riquelme2021scaling101010}. This sparsity-aware computation makes MoE architectures well-suited for balancing inference efficiency and model expressiveness \cite{szatkowski2024exploiting111111,liu2025sparsity121212}, enabling state-of-the-art performance across a range of complex AI tasks.

Despite these advantages, efficiently deploying MoE models in real environments remains challenging \cite{liu2025optimizing131313}. Unlike conventional single models that can be easily compressed or quantized for end deployment, MoE models consist of multiple dynamically invoked expert models, whose activation patterns vary per input and require efficient gating and routing mechanisms \cite{rajbhandari2022deepspeed141414}. This variability introduces new difficulties in system-level resource scheduling, model placement, and expert communication \cite{gale2023megablocks151515,cao2025moe161616}. On one hand, resource-constrained end devices struggle to support the intensive computational demands of high-capacity MoE backbones, especially when multiple expert paths are involved \cite{shen2024large171717,jin2025moe181818}. This results in degraded inference accuracy or throughput if only local computation is used \cite{li2025moe191919}. On the other hand, relying solely on the cloud introduces latency overhead and may lead to underutilization of local resources \cite{deshpande2024moesaic202020}. Moreover, cloud-only execution becomes highly sensitive to network fluctuations \cite{yu2024moesys212121}, making it difficult to guarantee consistent performance in latency-critical applications. These issues are further exacerbated by fluctuating network conditions, device heterogeneity, and dynamic workload patterns commonly encountered in real-world scenarios.


\par
To address these challenges, we propose \textit{EC$^2$MoE}, a novel adaptive framework that enables scalable and efficient MoE inference through end-cloud pipeline collaboration in heterogeneous distributed environments. \textit{EC$^2$MoE} integrates hardware-aware expert selection with coordinated execution across end and cloud, effectively mitigating resource constraints, network variability, and communication overhead. By jointly optimizing expert routing and inference scheduling, our framework achieves high throughput, low latency, and robust scalability under dynamic workloads. Our main contributions are as follows:

\par

\begin{itemize}
\item We propose a hardware-aware lightweight group gate network that efficiently adapts MoE expert selection to heterogeneous end-cloud environments. By incorporating a local expert selection mechanism based on device hardware characteristics and designing a lightweight group gate network module, it significantly reduces inference latency and routing overhead while maintaining expert selection quality.

\item We develop a collaborative end-cloud pipeline optimization mechanism tailored for scalable MoE inference. It integrates a low-rank compression-based encoder-decoder to reduce transmission costs and a route-aware heuristic pipeline scheduler that dynamically maps inference sub-tasks across end and cloud based on workload and communication patterns, maximizing overall throughput.

\item We evaluate the performance of the framework and compare it with mainstream baseline methods. Experimental results show that \textit{EC$^2$MoE} can increase throughput by 2.2$\times$ to 5.1$\times$ and reduce end-to-end latency by 53\% to 67\%, and without sacrificing accuracy. 

\end{itemize}

\section{Related Works}
\label{sec:formatting}

\subsection{Cloud-based MoE inference optimization} Cloud-based MoE inference optimization has been widely studied due to the abundant computing resources and scalability of cloud infrastructures \cite{hwang2024pre252525,hu2025brownoutserve262626}. Early works primarily focused on efficient expert routing and load balancing to minimize the communication and computation overhead during inference. For example, GShard \cite{lepikhin2020gshard222222} and Switch Transformer \cite{fedus2022switch232323} introduced sparse expert activation and simplified gating mechanisms to enable large-scale MoE training and inference in cloud environments. These methods significantly improved model scalability while controlling inference costs by activating only a subset of experts per input. Subsequent research further explored expert placement and communication optimization. Tutel \cite{hwang2023tutel555} and EfficientMoE \cite{10876795-272727} implemented expert parallelism strategies and expert sharding to minimize cross-device communication during MoE inference, enhancing throughput and reducing latency. In addition, model parallelism frameworks such as DeepSpeedMoE \cite{dai2024deepseekmoe242424} and Fsmoe \cite{pan2025fsmoe282828} provided system-level optimizations for cloud-based deployment by improving the scheduling of expert computation and network transfer. These works, however, often assume homogeneity and high bandwidth availability in cloud clusters, which may not generalize well to hybrid or end scenarios.

\par
\subsection{End-based MoE inference optimization}. 
Deploying MoE models on resource-constrained end devices faces limitations in terms of memory and computing power. To address this, previous studies have proposed various lightweight and dynamic management methods to improve inference efficiency. EdgeMoE \cite{10906629-292929} and D$^2$MoE \cite{wang2025d353535} achieve expert selection tailored to input and device status through sparse activation and dynamic routing strategies. Edge-MoE \cite{sarkar2023edge303030}, eMoE \cite{tairin2025emoe333333}, and Fate \cite{fang2025fate323232} focus on memory optimization, employing mechanisms such as expert sharing and cross-layer gating to reduce model overhead. AdapMoE \cite{zhong2024adapmoe313131} introduces a sensitivity-based gating mechanism to enable flexible precision control during inference. To address runtime resource fluctuations, the work \cite{11022729343434} proposes a dynamic expert replacement mechanism, while Flame \cite{lin2024flame363636} fully leverages MoE sparsity on FPGAs to achieve efficient hardware deployment. Although these methods significantly enhance end-side MoE inference capabilities, they remain constrained by model capacity and flexibility, making it challenging to meet the demands of high-precision tasks.

\begin{figure}[t!]
	\centering 
		\includegraphics[width=1\linewidth]{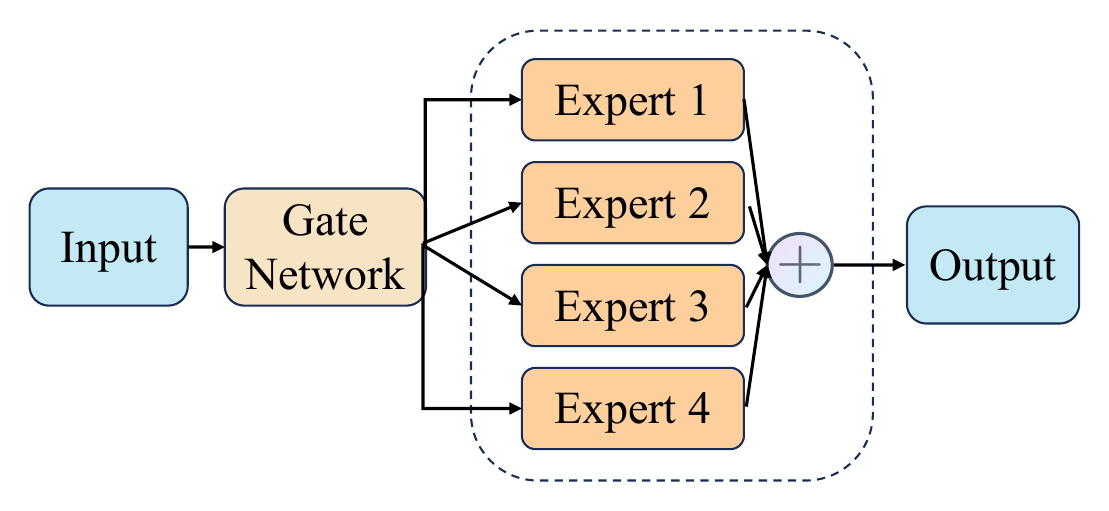}
	\caption{The illustration of MoE architecture, where a gating network dynamically selects a subset of expert networks to process each input. The selected expert outputs are then aggregated to form the final prediction.}
	\label{figure1}
\end{figure}

\section{Preliminary}

\subsection{Mixture-of-Experts}

MoE is an architecture that enhances model capacity and efficiency by dynamically activating submodels (experts), as shown in Figure~\ref{figure1}. Its core idea is to route input samples to a small number of relevant expert networks (such as feedforward layers) for processing, rather than having all parameters participate in the computation. By using a learnable gating mechanism (Gating Network) to select the top-k experts, MoE significantly increases the number of parameters while keeping the computational load close to that of a dense model.
The core of MoE lies in dynamically activating expert submodels through a gating network. Given an input $x$, its output $y$ can be represented as:

\begin{equation}
y=\sum_{i=1}^n G(x)_i \cdot E_i(x)
\end{equation}
where $E_i$ denotes the $i$ nd expert network (typically an independent feedforward layer), and $G(x)$ is the gating function output weight vector, satisfying $\sum_i G(x)_i=1$. A key advantage of MoE is its ability to perform effective pre-training with far fewer computational resources than dense models require. Compared to dense models, hybrid expert models typically achieve the same quality level faster. However, they may face challenges related to expert load balancing and cross-device communication overhead, especially in distributed deployments.

\subsection{End-Cloud Architecture}
The end-cloud architecture represents a distributed computing paradigm that strategically allocates computational workloads between resource-constrained end devices (e.g., smartphones and IoT sensors) and centralized cloud servers through collaborative execution mechanisms. In this architecture, end devices primarily perform latency-sensitive operations (e.g., real-time data preprocessing and lightweight model inference), while computationally intensive tasks (e.g., large-scale model training and complex inference) are offloaded to cloud servers. This hierarchical computation approach effectively addresses critical challenges in bandwidth utilization and end-to-end latency optimization. Particularly for MoE model deployment, the end-cloud architecture provides essential infrastructure support, as MoE systems inherently require distributed processing capabilities to handle their characteristic large-scale, diverse task distributions and dynamic expert routing mechanisms.

\section{Method}

\subsection{\textit{EC$^2$MoE} Overall Design}
\label{method}

\begin{figure}[t!]
	\centering 
		\includegraphics[width=1\linewidth]{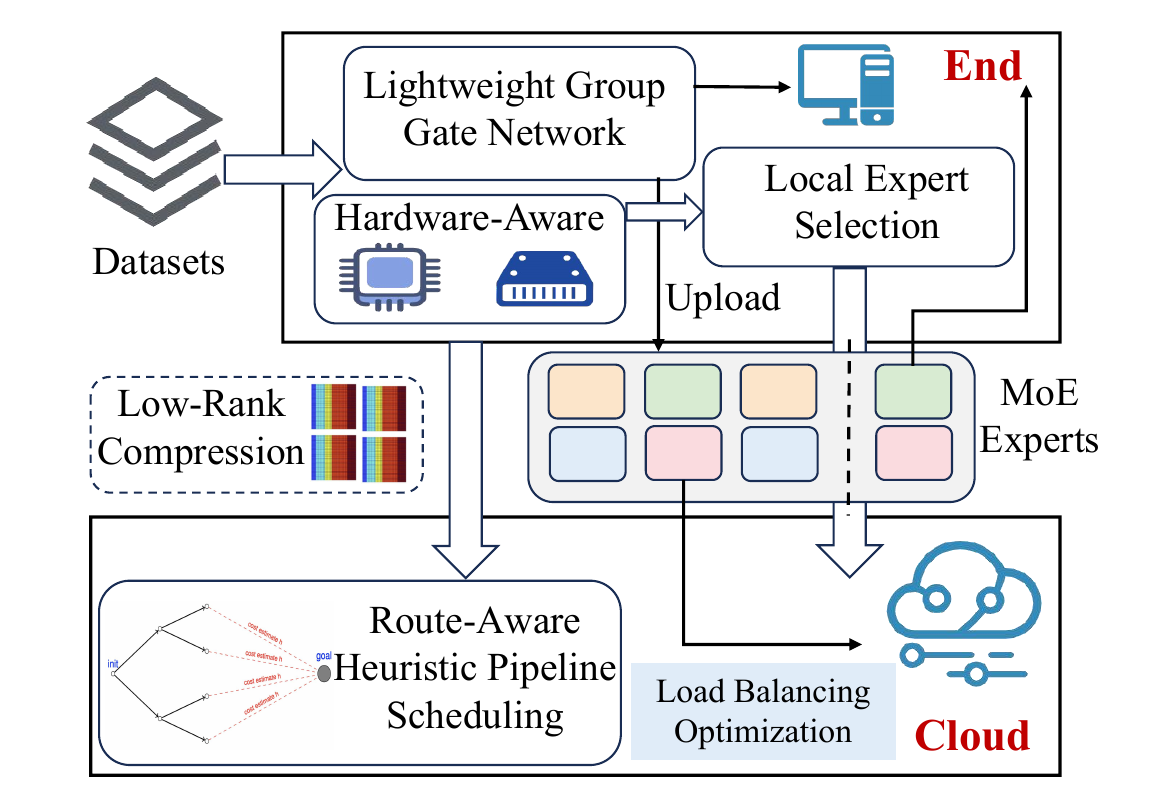}
	\caption{The overall design of the proposed \textit{EC$^2$MoE} framework.}
	\label{figure2}
\end{figure}

The overall workflow of \textit{EC$^2$MoE} consists of two key components: \textit{ (1) Hardware-Aware Lightweight Group Gate Network} (\textbf{HL-GGN}) and \textit{(2) Pipeline Optimization based on End-Cloud Collaboration} (\textbf{PO-ECC}). As shown in Figure~\ref{figure2}, first, the system performs hardware-aware local expert selection on the end, where candidate experts are filtered based on the device’s real-time resource profile to reduce unnecessary computation and transmission. Then, a lightweight group gate network fuses local gating signals with global expert routing to generate high-quality expert selection decisions with minimal overhead. Second, the selected inputs and routing information are passed through an end-cloud collaborative pipeline, where an encoder-decoder module based on low-rank compression reduces transmission latency and bandwidth cost, while a route-aware pipeline scheduler dynamically distributes inference stages across edge and cloud devices. This end-to-end collaborative workflow ensures optimal utilization of heterogeneous resources, improves system throughput, and maintains low latency under dynamic workloads and network conditions.

\subsection{Hardware-Aware Lightweight Group Gate Network}
To enable efficient expert selection in resource-constrained end environments, we propose the \textbf{HL-GGN}, which consists of two key components: (1) Hardware-Aware Local Expert Selection and (2) Lightweight Group Gate Network. This design ensures that only the most relevant experts are activated while minimizing computational overhead.

\subsubsection{Local Expert Selection} 
In heterogeneous end-cloud collaborative inference scenarios, end devices have limited computing power and cannot evaluate all expert models simultaneously during each inference. Therefore, we introduce a hardware-aware local expert selection mechanism to dynamically narrow down the set of expert candidates while maintaining the flexibility of the MoE architecture. Specifically, we define a hardware-aware function $H(\cdot)$, whose input is the real-time state vector of the device:

\begin{equation}
S_{\text {device }}=\left\{C_{c p u}, M_{\text {mem }}, P_{\text {power }}, B_{\text {bandwidth }}\right\}
\end{equation}
where $C_{cpu}$ denotes the currently available CPU resources, $M_{mem}$ represents the memory status, $P_{power}$ indicates the current battery level or power consumption limit, and $B_{bandwidth}$ signifies the network bandwidth condition. Based on the above states, the hardware-aware function can predict the current device's inference capability threshold $T_{capability}$:
\begin{equation}
T_{\text {capability }}=H\left(S_{\text {device }}\right)
\end{equation}

Subsequently, the computational complexity characteristics of the expert subnetwork are defined as vector $V_{expert}$, and by comparing the complexity characteristics of each expert network with the current device capability threshold, a subset of experts $\mathcal{E}_{local}$ that meet the local execution conditions is selected:

\begin{equation}
\mathcal{E}_{\text {local }}=\left\{e_i \mid f\left(V_{\text {expert }_i}, T_{\text {capability }}\right) \leq \epsilon, \forall e_i \in \mathcal{E}\right\}
\end{equation}

Here, $f(\cdot)$ is the complexity matching function, and $\epsilon$ is the set complexity tolerance threshold. Through this hardware-aware selection mechanism, we ensure that end devices only execute inference tasks within their capability range, while other high-complexity tasks are offloaded to the cloud, effectively balancing the real-time performance and resource consumption of inference.

\begin{figure}[t!]
	\centering 
		\includegraphics[width=1\linewidth]{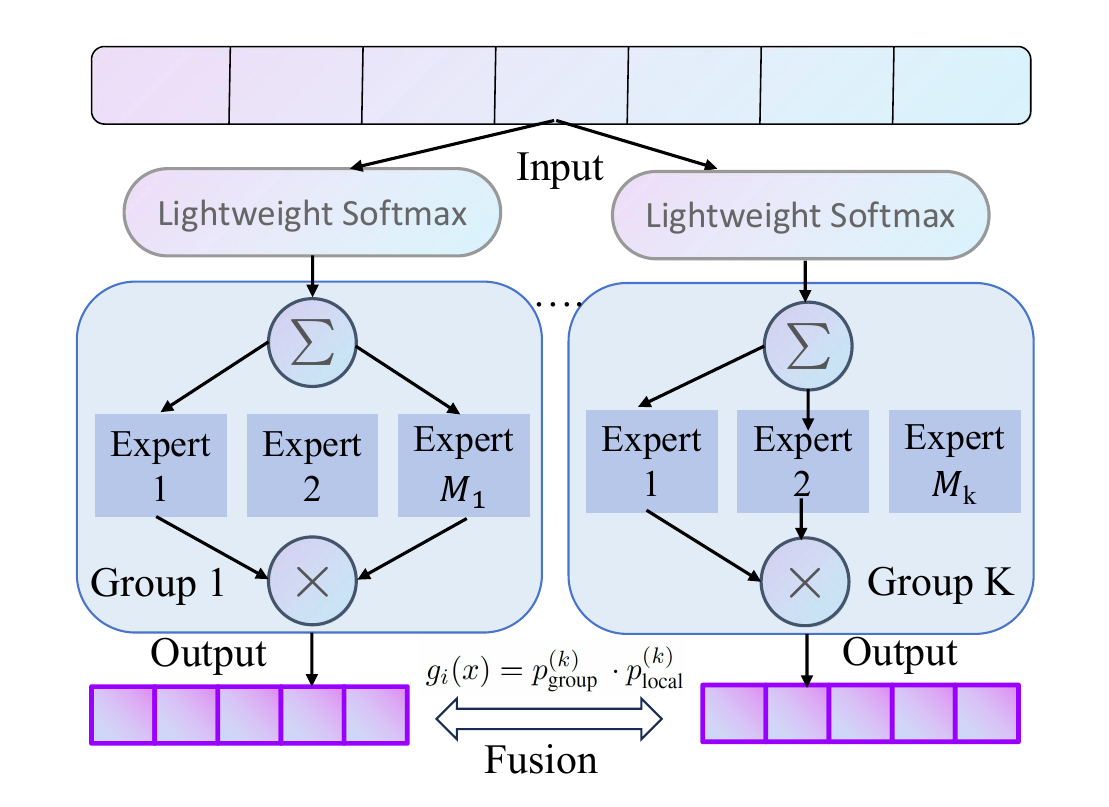}
	\caption{The overview of the lightweight group gate network.}
	\label{figure3}
\end{figure}

\subsubsection{Lightweight Group Gate Network}

Traditional gate networks generally calculate the selection probability of each expert network through a single fully connected layer \cite{ma2018modeling393939}. However, this design has the following shortcomings in end devices: (1) The weight matrix $W_g$ is linearly related to the number of experts and feature dimensions, making it difficult to deploy effectively in resource-constrained end devices \cite{petersen2022deep404040}. (2) High-dimensional matrix multiplication significantly increases inference latency.

To address the above issues, we first employ a group gate mechanism to divide the expert network into several groups, with each group sharing a separate gate subnetwork, thereby significantly reducing the number of parameters and computational cost of the gate network, as shown in Figure~\ref{figure3}. Specifically, the $M$ experts are divided into $K$ groups (each group contains $M_k$ experts, where $M = \sum_{k=1}^{K} M_k$), and each group independently learns a lightweight Softmax gate network:

\begin{equation}
\begin{aligned}
& g^{(k)}(x)=\operatorname{Softmax}\left(W_k x+b_k\right), \\
& W_k \in \mathbb{R}^{M_k \times d}, b_k \in \mathbb{R}^{M_k}
\end{aligned}
\end{equation}

Since the number of experts $M_k$ corresponding to each gate-controlled subnetwork is significantly smaller than the total number of experts $M$, the number of parameters and computational load for each subnetwork are greatly reduced. Furthermore, to achieve more flexible and accurate expert selection, we introduce a two-stage dynamic fusion strategy. In stage 1, a low-dimensional global Softmax gated network is used to quickly determine the overall contribution of each group of experts:

\begin{equation}
\begin{aligned}
& p_{\text {group }}=\operatorname{Softmax}\left(W_{\text {global }} x+b_{\text {global }}\right), \\
& W_{\text {global }} \in \mathbb{R}^{K \times d}, b_{\text {global }} \in \mathbb{R}^K
\end{aligned}
\end{equation}

Then, in Stage 2, within each expert group, the aforementioned lightweight grouped gating subnetwork is used to calculate the group-internal expert probability $p_{local}^{(k)}$. Finally, by multiplying and fusing the group-level and group-internal probabilities through the two-stage gating probabilities, the final expert selection probability is obtained:
\begin{equation}
g_i(x)=p_{\text {group }}^{(k)} \cdot p_{\text {local }, i}^{(k)}, \quad i \in \text { group } k
\end{equation}

Through this two-stage mechanism, it can dynamically focus on a small number of experts within highly correlated expert groups, further improving inference efficiency while maintaining good expert selection accuracy.

\subsection{Pipeline Optimization based on End-Cloud Collaboration}
\begin{figure}[t!]
	\centering 
		\includegraphics[width=1\linewidth]{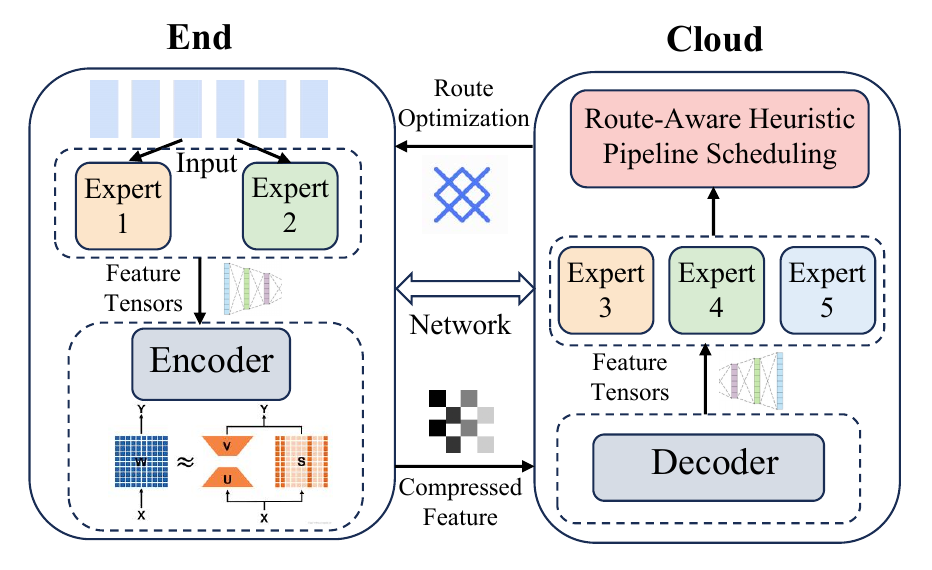}
	\caption{The overview of the pipeline optimization based on end-cloud collaboration.}
	\label{figure4}
\end{figure}

To achieve efficient and scalable MoE inference in an end-cloud collaborative environment, we propose the \textbf{PO-ECC} that combines low-rank compression for encoder-decoder and route-aware pipeline scheduling,  as shown in Figure~\ref{figure4}. It can ensure minimal communication overhead while maximizing computational parallelism.

\subsubsection{Encoder-Decoder based on Low-Rank Compression}
During the inference process involving frequent interactions between the end and the cloud, the transmission of a large number of intermediate feature tensors causes severe bandwidth pressure and communication latency \cite{hou2015sparse414141,idelbayev2020low424242}. To address this issue, we designed an encoder-decoder model based on low-rank compression to compress and reconstruct intermediate features, thereby reducing the communication burden between the end and the cloud.

Specifically, let the input feature tensor be $X \in \mathbb{R}^{h \times w \times c}$. We construct a lightweight compression module at the end to project it into a low-rank subspace, yielding the compressed representation $Z = U^\top X V$, where $U \in \mathbb{R}^{h \times r}$, $V \in \mathbb{R}^{w \times r}$ is an adaptively optimized low-rank projection matrix, and $r \ll \min(h, w)$ denotes the compression dimension.

On the cloud, the decoder module uses the corresponding inverse projection matrices $\hat{U}, \hat{V}$ to reconstruct the original feature tensor $\hat{X} = \hat{U} Z \hat{V}^\top$. To mitigate the accuracy degradation caused by reconstruction errors, we adopt an end-cloud joint training approach to minimize the following objective function:

\begin{equation}
\mathcal{L}_{\text {rec }}=\|X-\hat{X}\|_2^2+\lambda \cdot \mathcal{L}_{\text {task }}(\hat{X})
\end{equation}
where $\mathcal{L}_{task}$ is the task loss function, and $\lambda$ controls the trade-off between compression and task accuracy. This method effectively compresses communication data while maintaining accuracy, significantly reduces transmission latency between the end and cloud, and provides a solid foundation for subsequent pipeline scheduling.

\subsubsection{Route-Aware Heuristic Pipeline Scheduling }

To further improve the overall processing efficiency of pipeline tasks in an end-to-end cloud collaboration environment, we propose a route-aware heuristic pipeline scheduling algorithm. Specifically, the pipeline scheduling problem can be abstracted as follows: given a set of tasks $\mathcal{T} = {t_1, t_2, ..., t_N}$, each task can be assigned to either end or cloud processing, and different tasks have different computational complexities $C(t_i)$ and communication costs $Comm(t_i)$. The heuristic optimization objective is defined as:

\begin{equation}
\min \sum_{i=1}^N\left[\alpha \cdot \operatorname{ExecTime}\left(t_i\right)+(1-\alpha) \cdot \operatorname{Comm}\left(t_i\right)\right]
\end{equation}
where $ExecTime(t_i)$ denotes the expected execution time of task $t_i$ when processed on the end or in the cloud. $Comm(t_i)$ represents the communication cost of task transmission. The parameter $\alpha$ is used to adjust the trade-off between computation and communication. To effectively solve the above optimization problem, we use a heuristic algorithm based on greedy selection for pipeline scheduling. First, we calculate the priority $P(t_i)$ of each task:
\begin{equation}
P\left(t_i\right)=\frac{C\left(t_i\right)}{\operatorname{Comm}\left(t_i\right)+\epsilon}
\end{equation}
where $\epsilon$ is a small constant to prevent division by zero. Tasks with higher priorities tend to be executed locally to reduce communication costs.

Then, based on the current end device load $Load_{end}$, cloud load $Load_{cloud}$, and the priority $P(t_i)$ of each task, the heuristic algorithm decides the execution location of the task:

\begin{equation}
\text { Location }\left(t_i\right)= \begin{cases}\text { End, } & \text { if Load }{ }_{\text {end }}+C\left(t_i\right) \leq T_{\text {end }} \\ & \text { and } P\left(t_i\right) \geq \beta \\ \text { Cloud, } & \text { otherwise }\end{cases}
\end{equation}
where $T_{end}$ is the maximum tolerable computation threshold for end devices, and $\beta$ is the priority threshold for local task execution. This heuristic rule enables rapid task allocation decisions and reduces task waiting latencys. Through the above routing-aware heuristic pipeline scheduling algorithm, it can efficiently respond to the dynamic nature of heterogeneous communication environments and improve the overall inference efficiency of the end-to-cloud pipeline.

\section{Performance Evaluation}  \label{Performance Evaluation}

\subsection{Evaluation Setup}
\subsubsection{Datasets and Implementation Details} 

To evaluate the effectiveness of the proposed framework, we conduct experiments on two widely used benchmark datasets: the GLUE \cite{GLUE373737} and the SQuAD \cite{rajpurkar2016squad383838}. We adopt the Switch Transformer as our base MoE model. The MoE inference pipeline is implemented using PyTorch, with custom modules for expert partitioning, dynamic routing, and end-cloud collaboration. Hardware simulations include an Intel Xeon Silver 4214R CPU for end devices and two NVIDIA A100 GPUs for cloud servers, reflecting a realistic heterogeneous compute environment. To simulate real network conditions, we use Linux Traffic Control to introduce network bandwidth settings. The communication between end and cloud is set under a 300 Mbps network with a 20\% fluctuation to reflect dynamic bandwidth conditions.

\par

\textbf{Baseline Methods}. We compare our solution with the following four baseline methods. (1) \textbf{BrownoutServe:} This is a cloud-based MoE inference service framework that reduces the number of expert visits and lowers inference latency by introducing a joint expert mechanism. (2) \textbf{EdgeMoE:} This is an MoE inference engine based on end devices that achieves memory and computational efficiency by dividing the model into a storage hierarchy.

\par

\textbf{Evaluation Metrics:} Three key metrics are employed to comprehensively assess the performance of the proposed framework: accuracy, end-to-end latency, and throughput. Accuracy is used to evaluate the correctness of model predictions and ensure that efficiency gains do not compromise model performance. End-to-end latency measures the total time from input reception at the end to final output generation, capturing the real-time responsiveness of the end-cloud collaborative inference pipeline. Throughput quantifies the number of inference tasks completed per unit time, reflecting the system’s overall processing efficiency.

\textbf{Parameter Selection:} The number of experts in the MoE models is set to 8, 16, 32, and 64, enabling the assessment of the system’s scalability and its adaptability to models of increasing complexity. The input sequence length is fixed at 256 tokens, which aligns with common benchmarks in language modeling tasks and ensures a consistent inference context across all experiments. A uniform batch size of 4 is used to balance computational efficiency and memory usage, especially under resource-constrained end settings. To reduce routing complexity and maintain inference efficiency, the gating mechanism adopts a Top-1 expert selection policy, where only the expert with the highest activation score is selected for each input, thereby minimizing redundant computation. Furthermore, to reflect hardware-aware constraints on the end side, a local expert selection mechanism is employed, with a selection cap that restricts the candidate expert set to at most 40\% of the total experts. This constraint ensures that only a small subset of experts is evaluated on the end device, significantly reducing computation and communication overhead while retaining the flexibility and performance benefits of sparse expert activation.

\subsection{End-to-End Results}

\subsubsection{Accuracy Comparison}

\begin{table*}[th]
\centering
\caption{The accuracy comparison results under the different datasets.}
\label{tab2}
\setlength{\tabcolsep}{2mm}{
\begin{tabular}{cccccccc}
\toprule
\multirow{2}{*}{Switch-Base} & \multicolumn{3}{c}{GLUE} & \multicolumn{3}{c}{SQuAD} \\
\cmidrule(lr){2-4} \cmidrule(lr){5-7}
 &BrownoutServe &EdgeMoE & \textit{EC$^2$MoE} & BrownoutServe &EdgeMoE & \textit{EC$^2$MoE} \\
\midrule
8-experts & 81.2 & 77.6 &  \textbf{81.6} & 82.2 & 78.4 & \textbf{82.5} \\
16-experts & 81.7 & 77.9 & \textbf{82.3} & 82.1 & 77.9 & \textbf{82.3} \\
32-experts & 80.3 & 76.2 &  \textbf{80.5} & 82.7 & 78.1 & \textbf{83.1} \\
64-experts & 80.8 & 77.8 & \textbf{81.4} & 82.3 & 77.3 & \textbf{82.6} \\
\bottomrule
\end{tabular}}
\end{table*}

The experimental results are shown in Table 1. The \textit{EC$^2$MoE} method proposed in this paper achieves the highest accuracy on both datasets. Compared with EdgeMoE, the average improvement reaches 4.07\% and 4.1\%. More importantly, the proposed method not only outperforms EdgeMoE but also surpasses the purely cloud-based method BrownoutServe. This is because cloud-based methods often assume stable network transmission and rely on global expert activation. However, in real-world end-cloud environments, high latency and bandwidth fluctuations may limit expert selection for certain inputs, thereby affecting inference integrity and final accuracy. This paper's method employs end-cloud joint gating and pipeline-level scheduling optimization to perform local selection and feature compression as inputs flow through the end, transmitting more representative information to the cloud and thereby reducing accuracy losses caused by network uncertainty. In contrast, EdgeMoE methods generally have lower accuracy rates due to the limited computing power and storage of end devices, which prevent the evaluation of the entire expert set, resulting in a limited number of activated experts and significant bias in the selection results. Although EdgeMoE avoids network transmission overhead locally, its disadvantages in expert diversity and model capacity utilization make it difficult to maintain the same inference accuracy as cloud-based methods in complex tasks.

\subsubsection{Throughput Comparison}

\begin{figure}[th]
	\centering 
    \subfloat[ GLUE]{
    \label{Fig.sub.5.1}
    \includegraphics[width=0.21\textwidth]{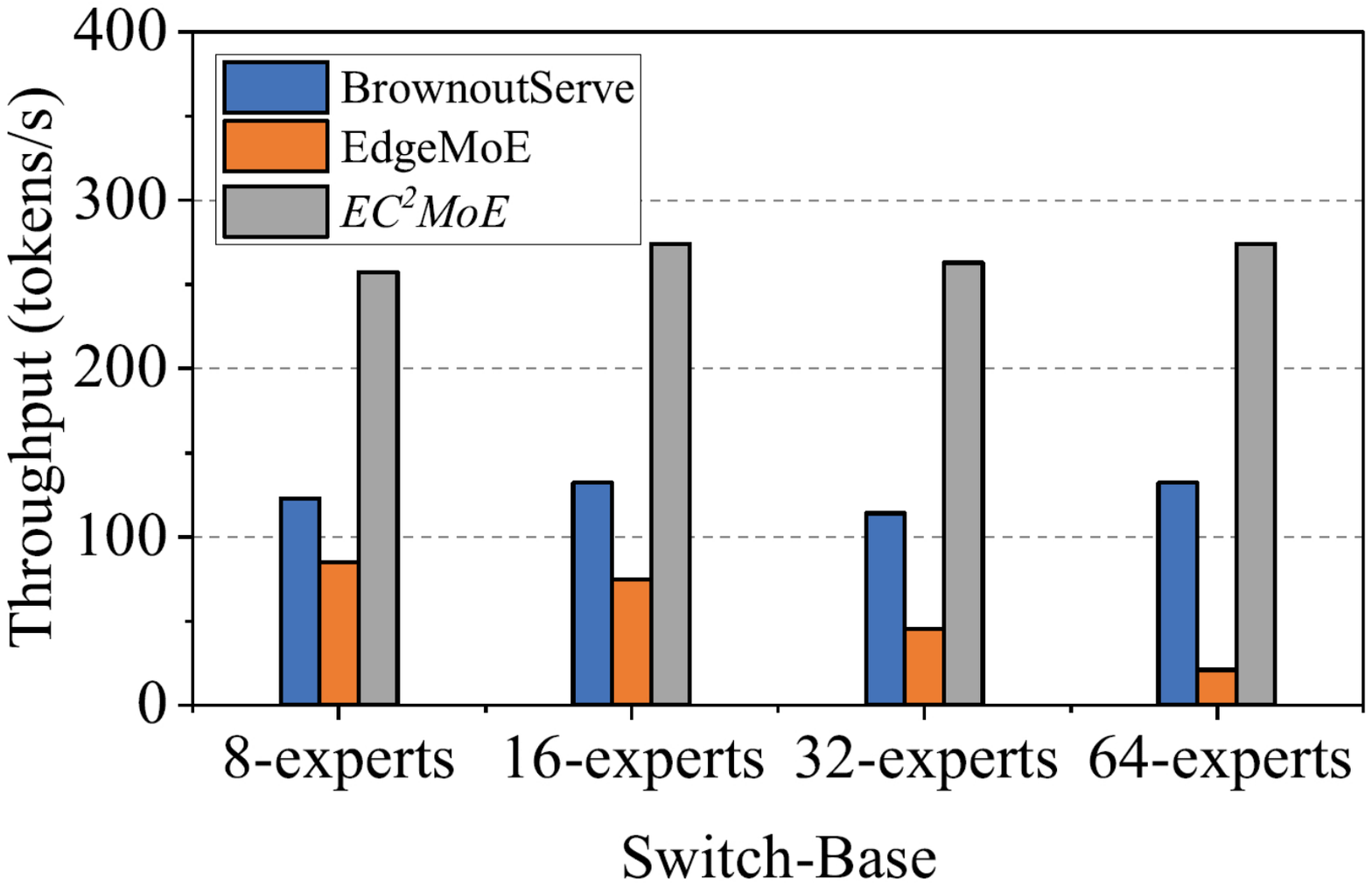}}
    \subfloat[SQuAD]{
    \label{Fig.sub.5.2}
    \includegraphics[width=0.21\textwidth]{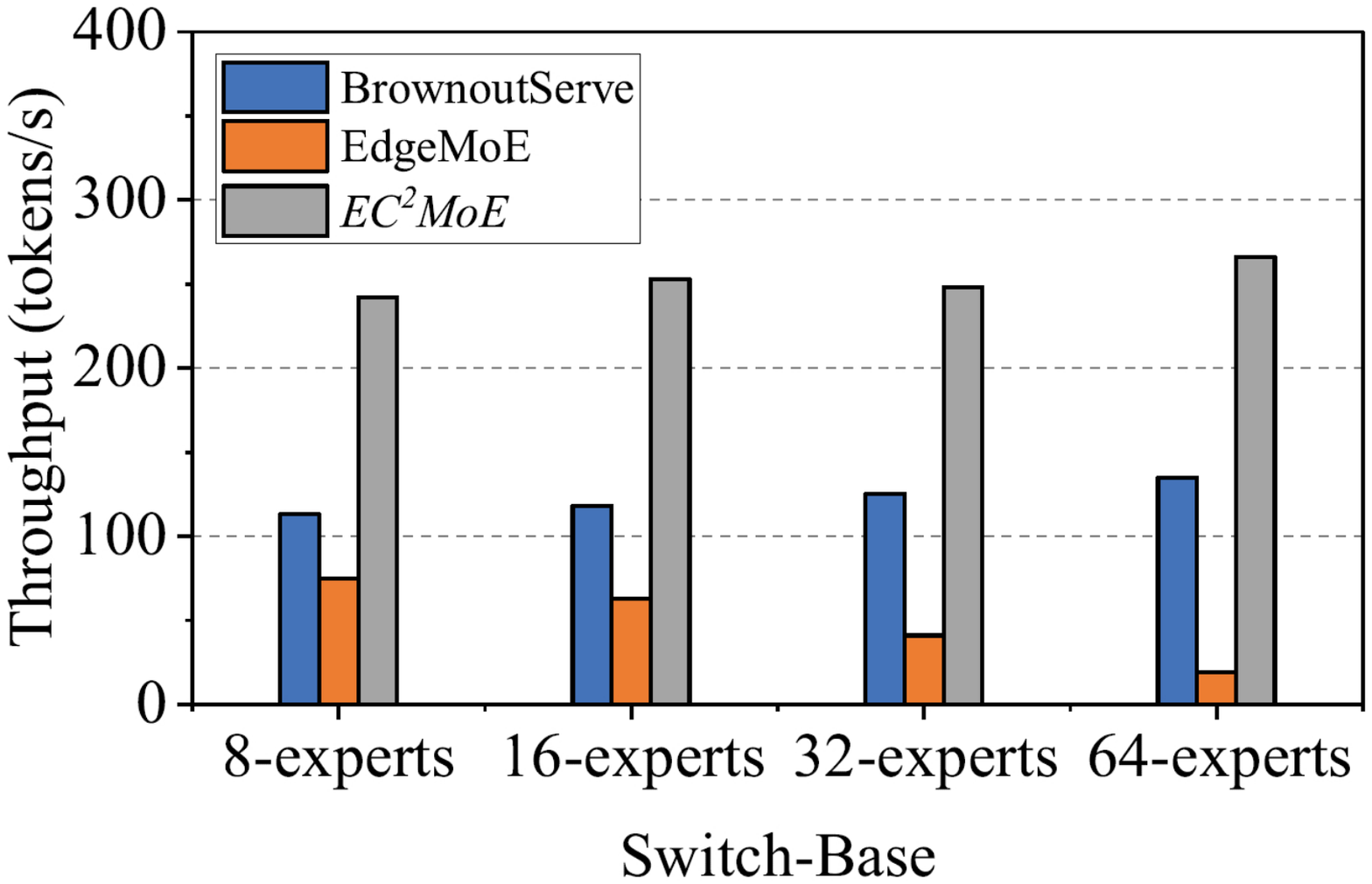}}
    \caption{The throughput comparison of different methods under different numbers of experts.}
	\label{figure5}
\end{figure}

The experimental results show that \textit{EC$^2$MoE} demonstrates significant throughput advantages across all expert scales, as shown in Figure~\ref{figure5}. Specifically, compared to BrownoutServe, \textit{EC$^2$MoE} achieves an average throughput improvement of over 2.2$\times$ at expert settings of 8, 16, 32, and 64. Compared to EdgeMoE, the average improvement reaches over 5.1$\times$. The fundamental reason for this improvement lies in this paper introduces a routing-aware end-cloud asynchronous scheduling mechanism that fully leverages the heterogeneous parallel capabilities of the end and cloud to maximize inference process overlap and throughput. And hardware-aware expert selection strategies are employed to asynchronously transmit only critical information to the cloud, thereby significantly reducing communication overhead and redundant computational load. In contrast, BrownoutServe processes most of the inference process in the cloud, which, despite its strong computing power, is limited by network latency and bandwidth fluctuations and cannot efficiently process a large number of requests in parallel. EdgeMoE, on the other hand, relies entirely on the local computing power of terminal devices, which have a short inference path but limited processing capacity, especially when the number of experts increases, leading to a significant bottleneck in resources and a rapid decline in throughput.

\subsubsection{Latency Comparison}
\begin{figure}[th]
	\centering 
    \subfloat[ GLUE]{
    \label{Fig.sub.6.1}
    \includegraphics[width=0.21\textwidth]{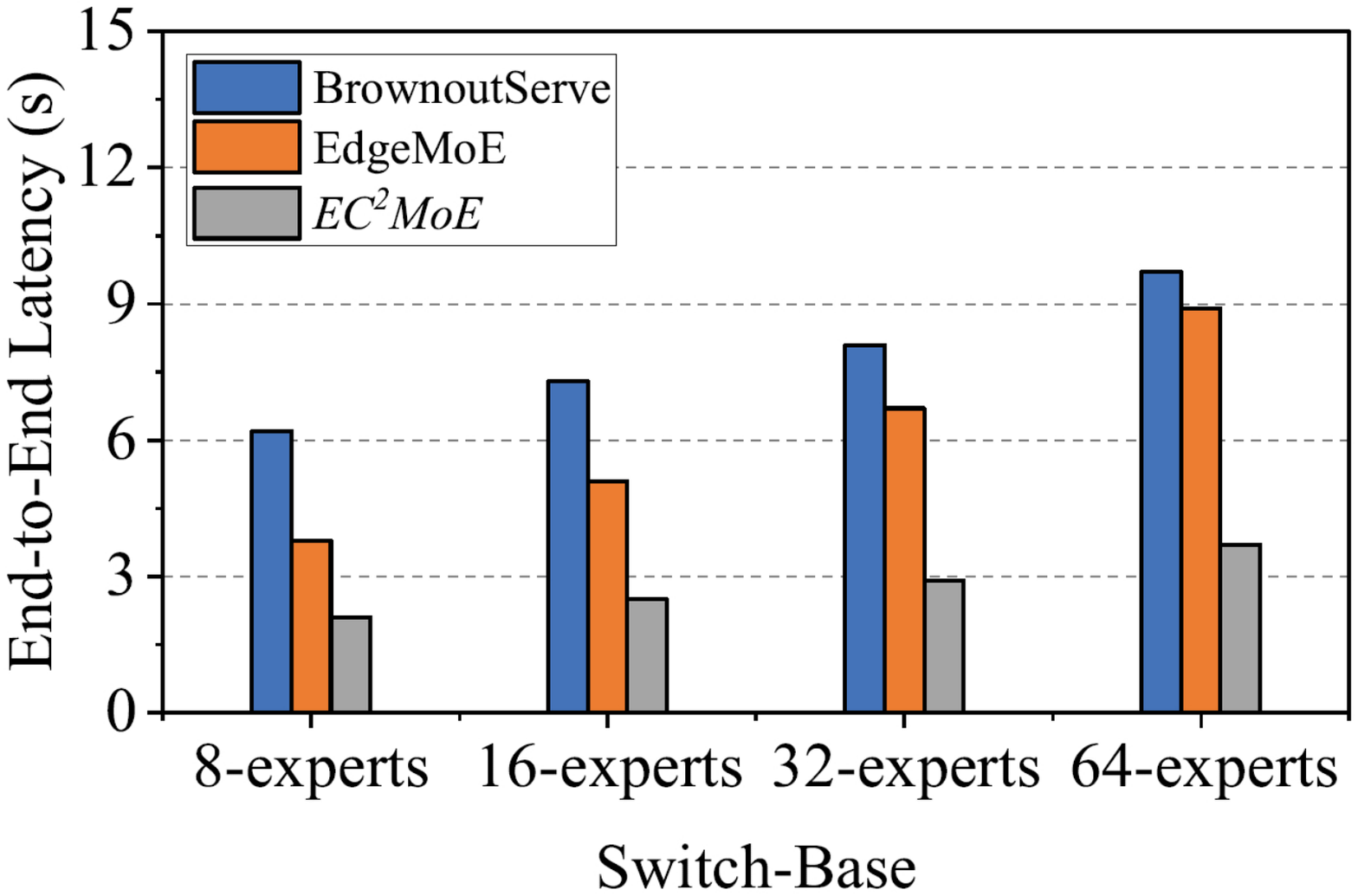}}
    \subfloat[SQuAD]{
    \label{Fig.sub.6.2}
    \includegraphics[width=0.21\textwidth]{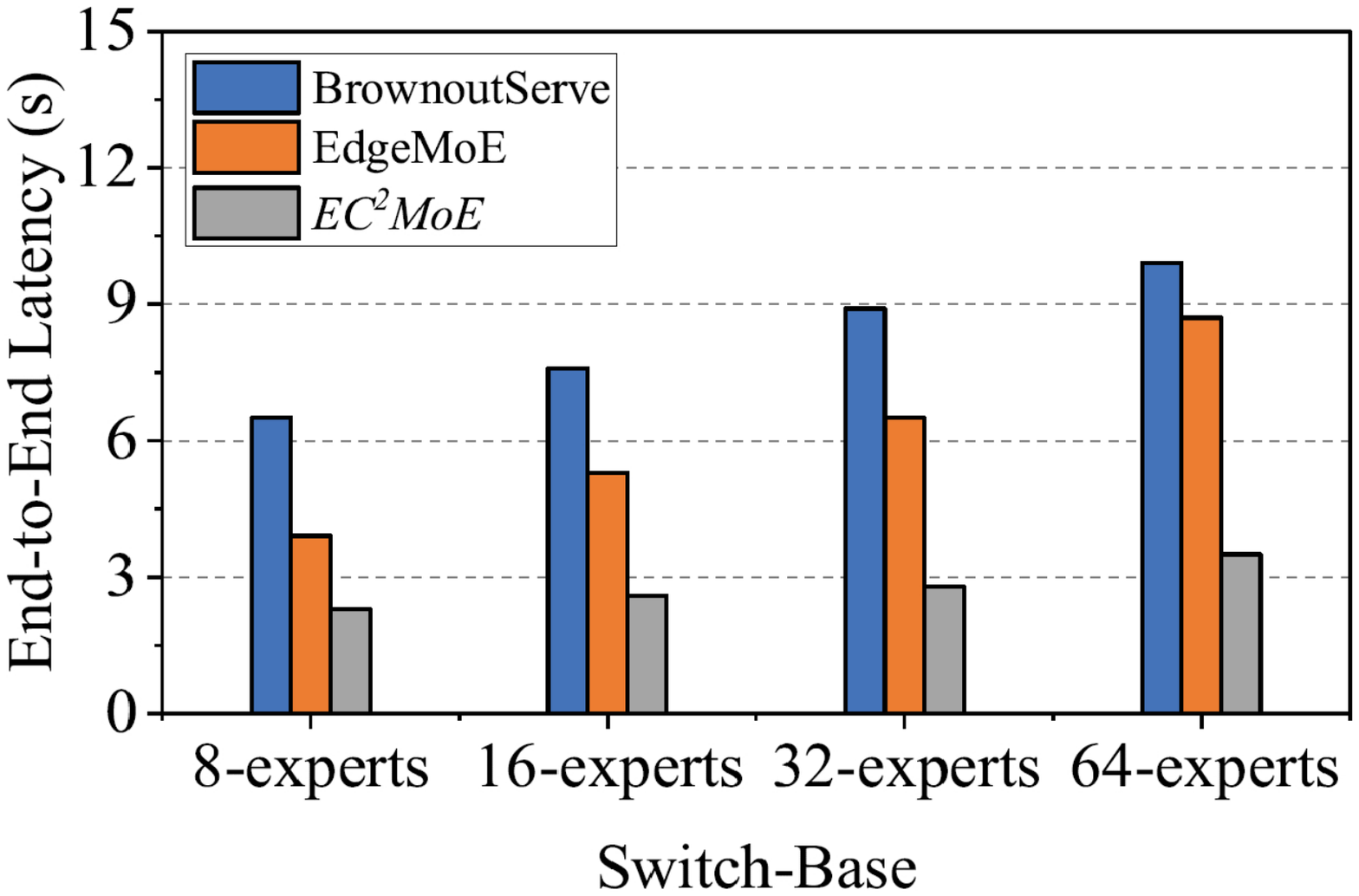}}
    \caption{The latency comparison of different methods under different numbers of experts.}
	\label{figure6}
\end{figure}

As shown in Figure~\ref{figure6}, the experimental results show that our proposed method also has significant advantages in terms of latency performance. Compared to the BrownoutServe method, \textit{EC$^2$MoE} can reduce the average latency by 67\%. Compared to the EdgeMoE method, the average latency is reduced by 53\%. \textit{EC$^2$MoE} can be achieved through a routing-aware task scheduling mechanism, so that the end-side and cloud-side can work in parallel, thereby significantly reducing overall transmission and inference latency. Specifically, the high latency of the BrownoutServe method mainly stems from its high dependence on network transmission. In actual environments, network instability and latency fluctuations can easily cause transmission bottlenecks, thereby delaying the overall inference process. Especially in scenarios where the number of experts increases, the cost of cloud resource scheduling rises significantly, further increasing the system response time. While the EdgeMoE method avoids network transmission latency, it is constrained by the computational power of terminal devices, resulting in limited processing capabilities during expert selection and model execution. Especially when dealing with high-capacity MoE structures, computational bottlenecks can easily form.

\subsection{Scalable Analysis}

\subsubsection{Task Load Changes}
\begin{figure}[th]
	\centering 
    \subfloat[]{
    \label{Fig.sub.7.1}
    \includegraphics[width=0.18\textwidth]{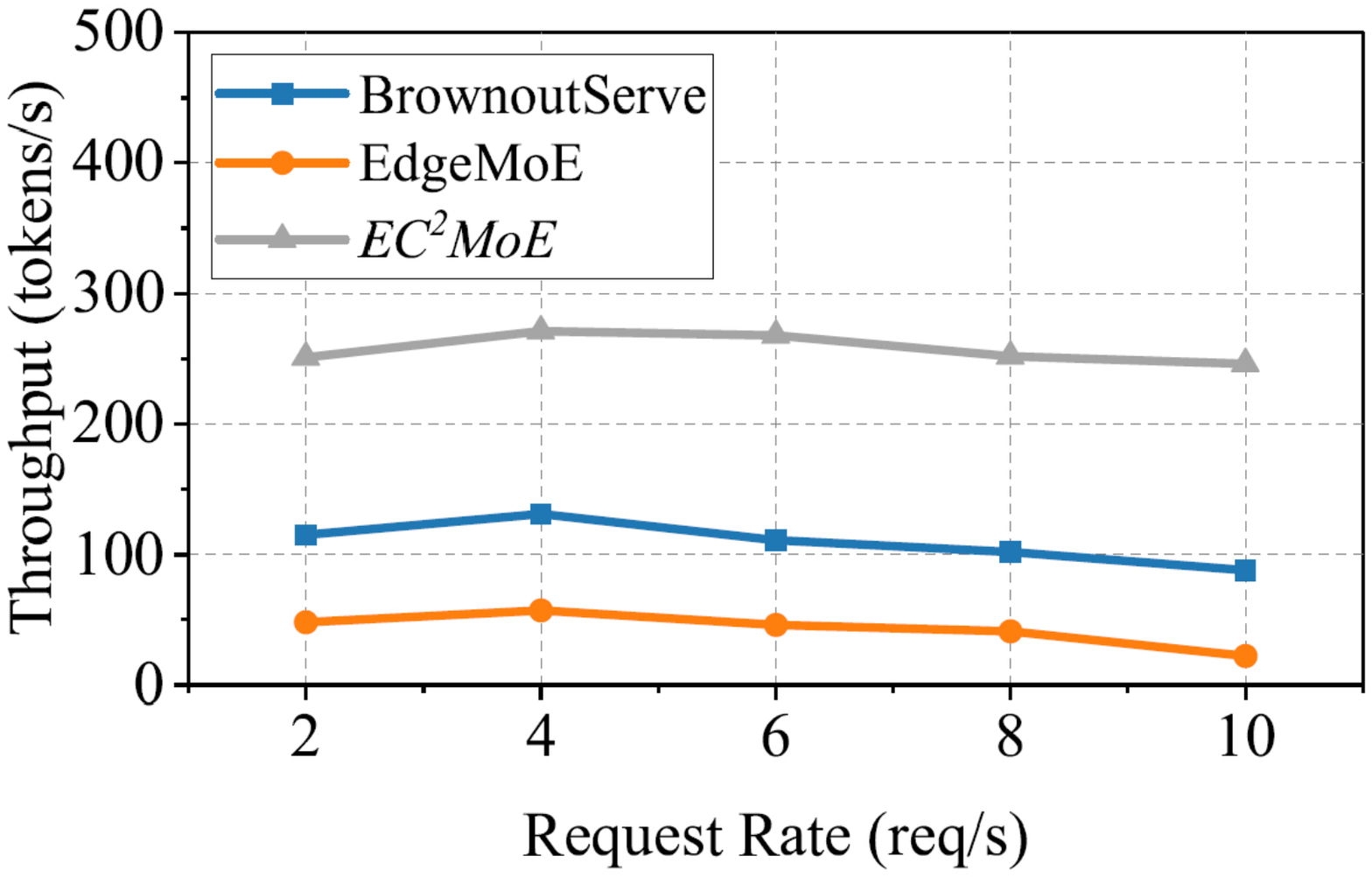}}
    \subfloat[]{
    \label{Fig.sub.7.2}
    \includegraphics[width=0.18\textwidth]{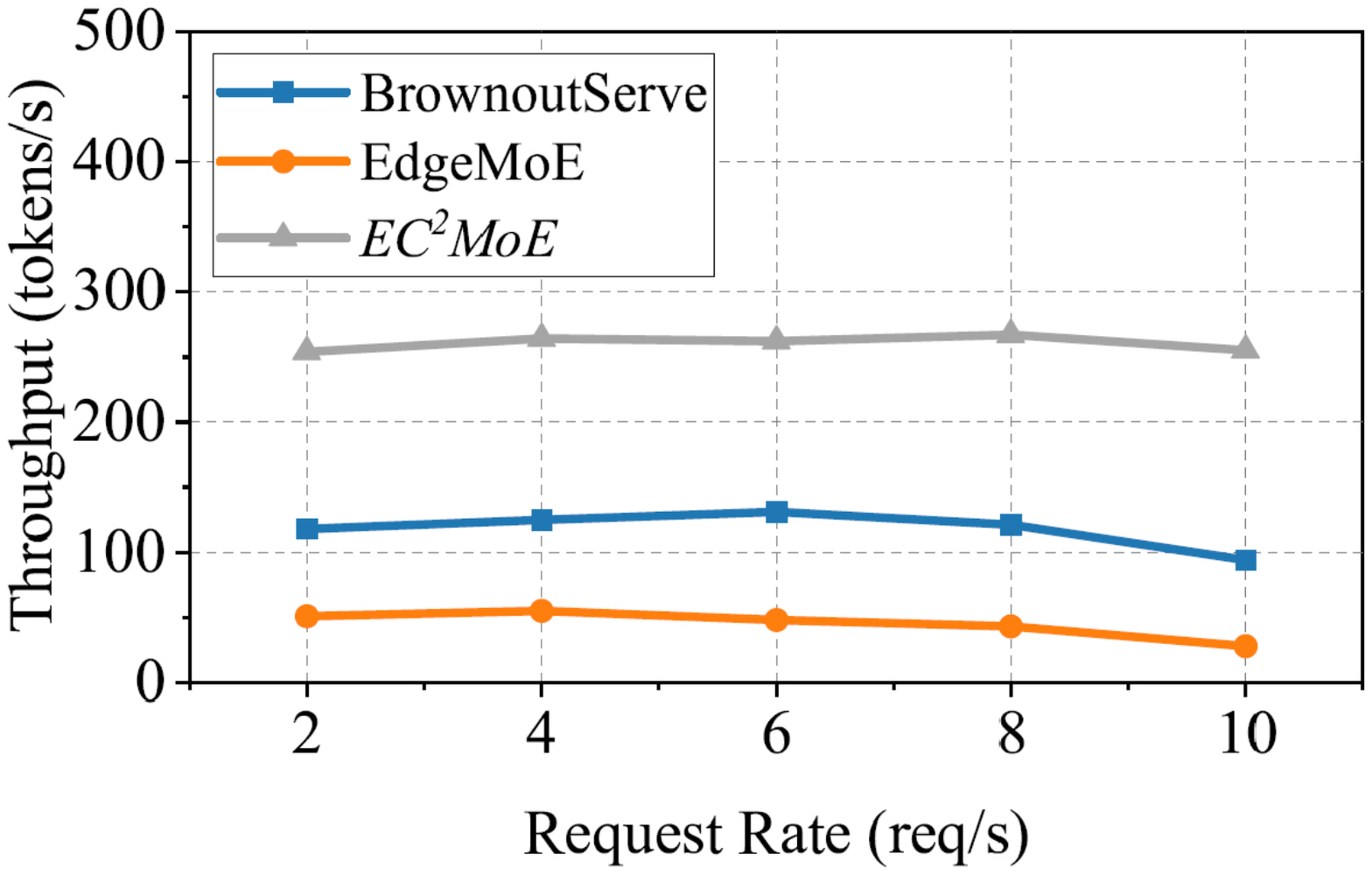}}
    
       \subfloat[]{
    \label{Fig.sub.7.3}
    \includegraphics[width=0.18\textwidth]{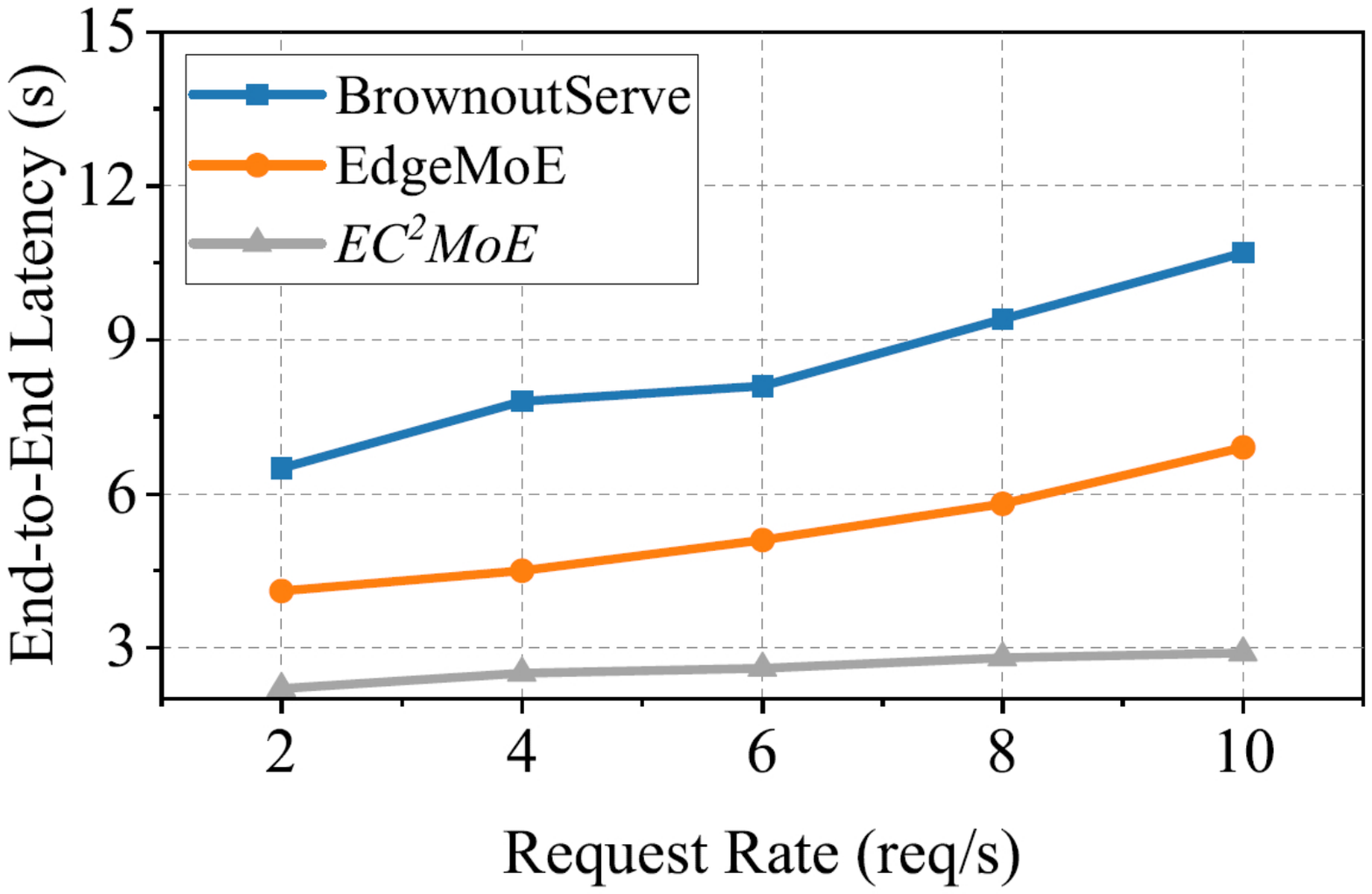}}
       \subfloat[]{
    \label{Fig.sub.7.4}
    \includegraphics[width=0.18\textwidth]{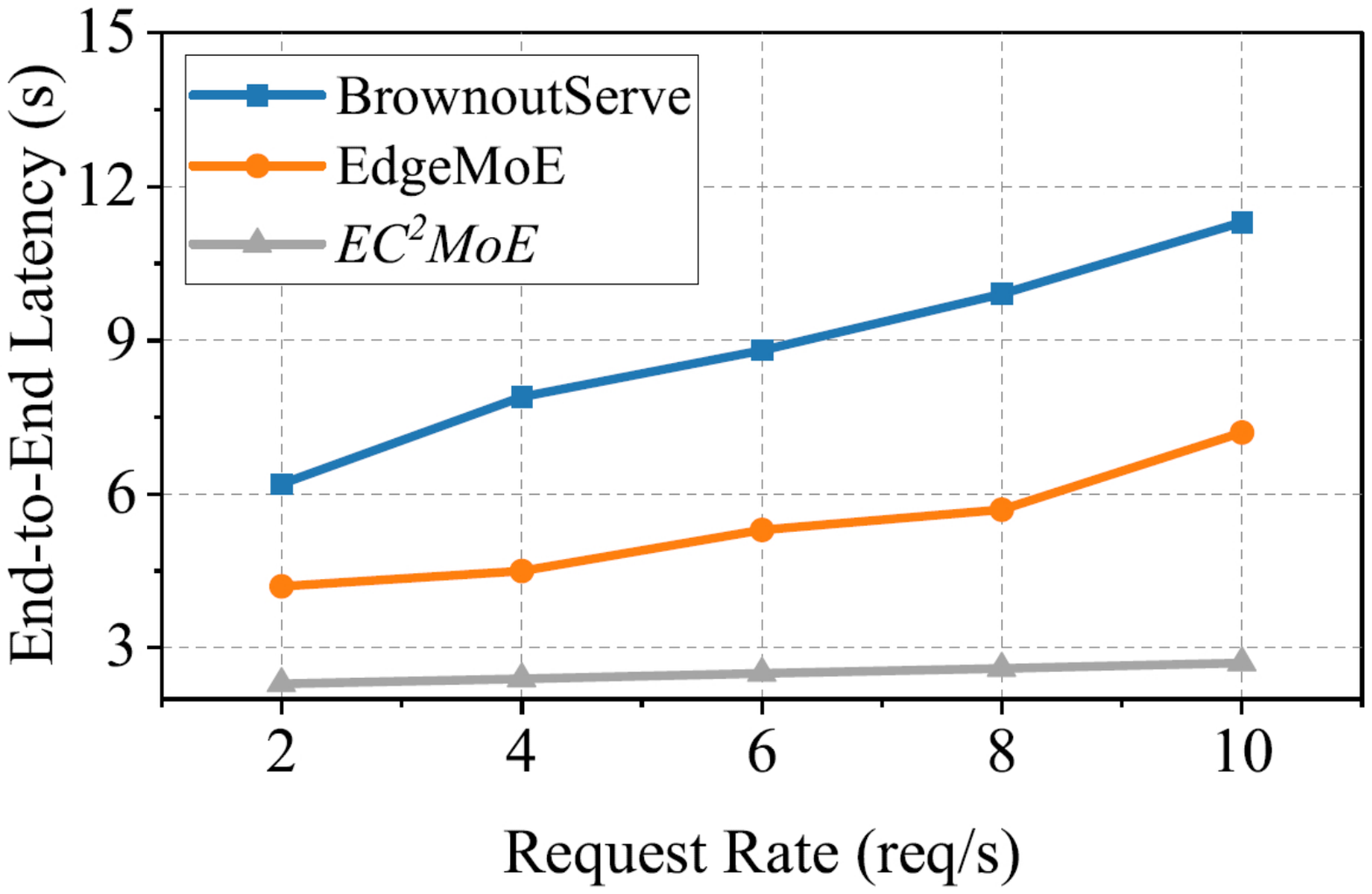}}
    \caption{The test results under different request rates. (a) and (b) are throughput results under the GLUE and SQuAD datasets. (c) and (d) are latency results under the GLUE and SQuAD datasets.}
	\label{figure7}
\end{figure}

To assess the scalability of \textit{EC$^2$MoE} under different reasoning load intensities, we simulated five request rates (Request Rate = 2, 4, 6, 8, 10 req/s) and statistically analyzed the average performance of throughput and end-to-end latency under different expert quantity settings, as shown in Figure~\ref{figure7}. It can be seen that as the request rate gradually increases, the method proposed in this paper demonstrates superior linear scalability in throughput, maintaining steady growth even under high loads. At the same time, the increase in latency is significantly lower than that of the comparison methods. In contrast, BrownoutServe is constrained by cloud processing resources and network congestion at high request rates, resulting in saturated throughput growth and a sharp increase in latency. EdgeMoE, on the other hand, quickly reaches a processing bottleneck in high-concurrency scenarios due to the limited computing power of end devices. The end-to-cloud pipeline collaboration mechanism adopted in this paper achieves high matching between processing capacity and input intensity by reasonably splitting the inference path and dynamically distributing task loads.

\subsubsection{Dynamic Network Environment}
\begin{figure}[th]
	\centering 
    \subfloat[]{
    \label{Fig.sub.11.1}
    \includegraphics[width=0.18\textwidth]{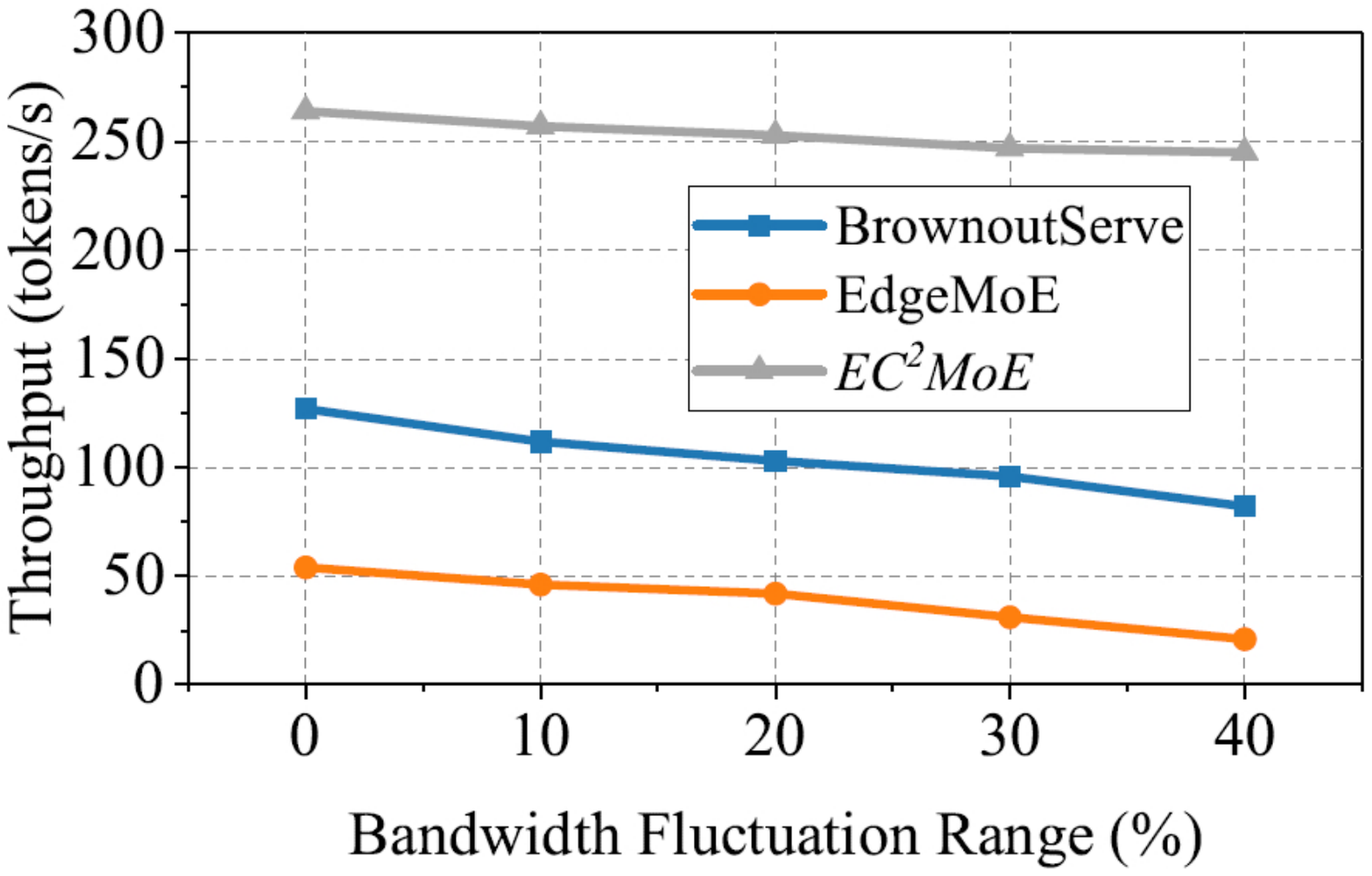}}
    \subfloat[]{
    \label{Fig.sub.11.2}
    \includegraphics[width=0.18\textwidth]{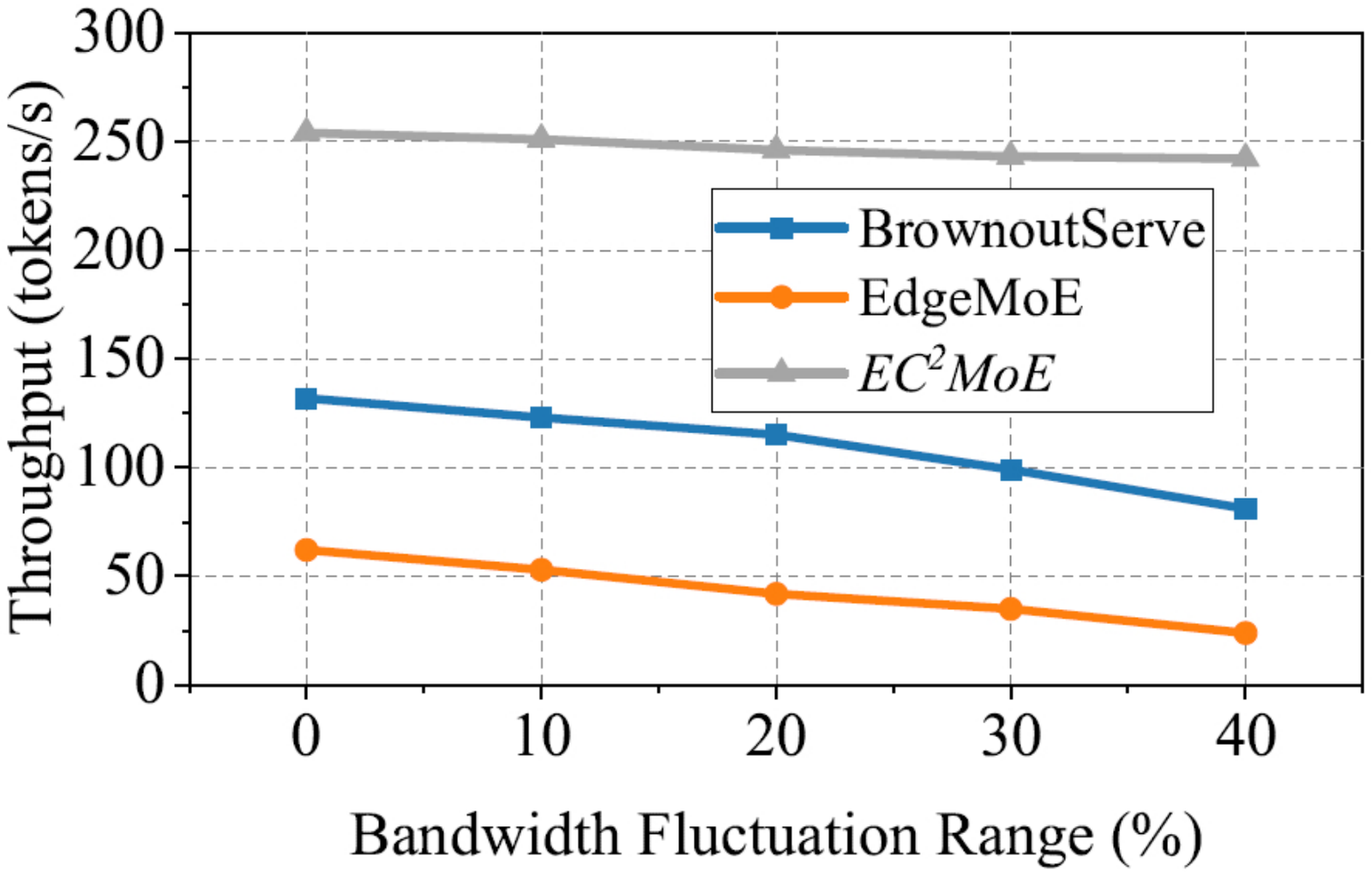}}
    
       \subfloat[]{
    \label{Fig.sub.11.3}
    \includegraphics[width=0.18\textwidth]{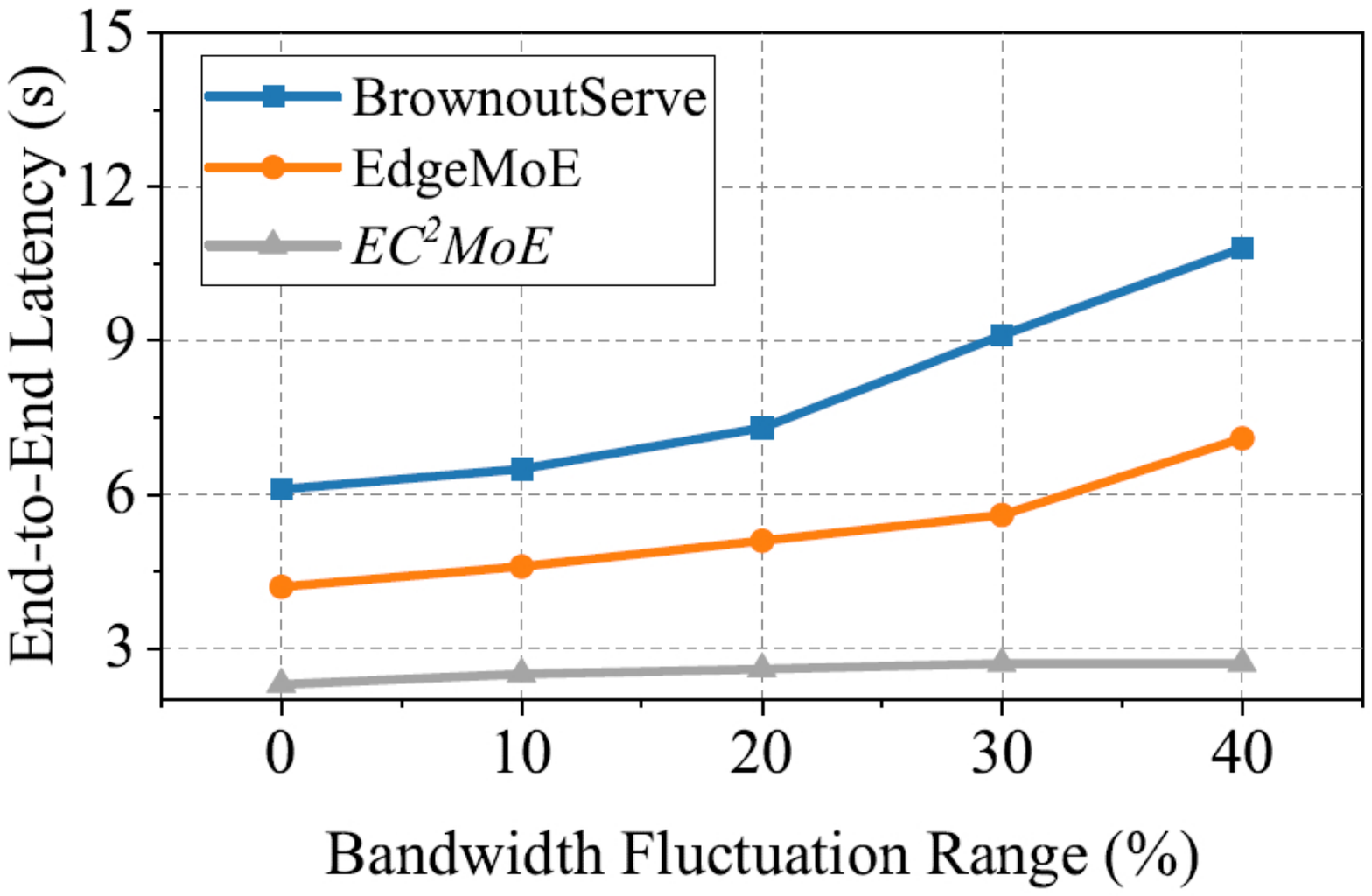}}
       \subfloat[]{
    \label{Fig.sub.11.4}
    \includegraphics[width=0.18\textwidth]{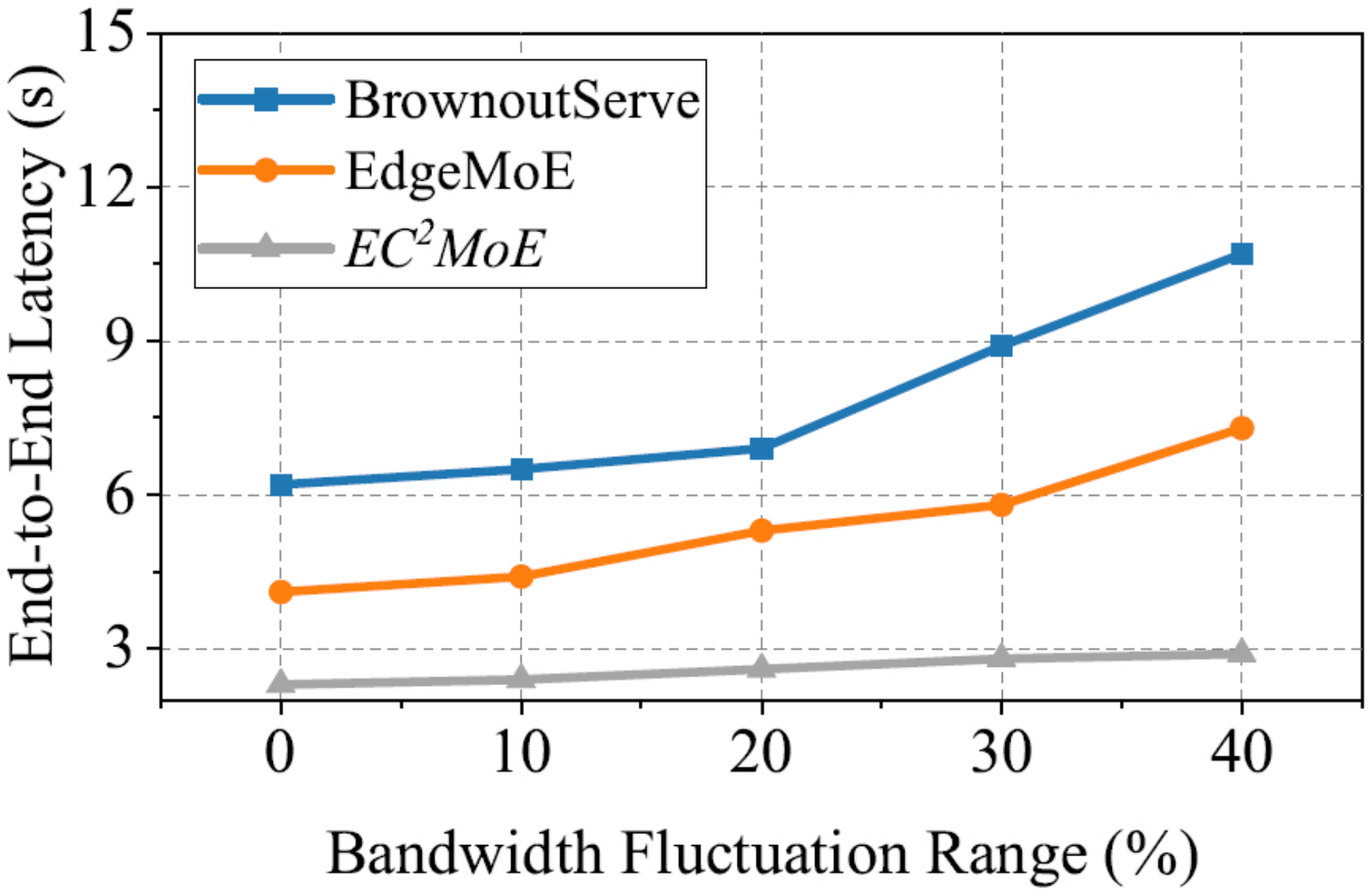}}
    \caption{The test results under different bandwidth fluctuation. (a) and (b) are throughput results under the GLUE and SQuAD datasets. (c) and (d) are latency results under the GLUE and SQuAD datasets.}
	\label{figure8}
\end{figure}

To further validate the scalability of \textit{EC$^2$MoE} in dynamic network environments, we set five network bandwidth fluctuation ranges (Bandwidth Fluctuation = 0\%, 10\%, 20\%, 30\%, 40\%). The experimental results are shown in Figure~\ref{figure8}. Even under conditions of increased network fluctuations, \textit{EC$^2$MoE} maintains stable throughput and latency, significantly outperforming BrownoutServe and EdgeMoE. BrownoutServe is highly sensitive to bandwidth fluctuations, and its cloud-based full-path dependency leads to significant latency increases and throughput collapse in high-jitter environments. While EdgeMoE does not rely on the network, its local inference capabilities cannot handle complex inference tasks with a high number of experts, resulting in limited throughput. In contrast, \textit{EC$^2$MoE} effectively reduces dependence on bandwidth stability through its strategy of local expert selection and asynchronous transmission, demonstrating excellent scalability and robustness even under uncertain network conditions.

\subsection{Ablation Studies}
To validate the contribution of each core module to overall performance, we removed two key design components: (1) HL-GGN and (2) PO-ECC. While keeping all other settings unchanged, the system's accuracy, throughput, and latency changes were evaluated on the GLUE and SQuAD datasets. Experimental results show that after removing HL-GGN, the system could not adequately adapt to the resource state of the terminal device during the expert selection phase, leading to a decrease in expert activation accuracy, an average reduction in overall accuracy of 2.1\%, and an increase in latency of approximately 23\%. Removing PO-ECC prevents the inference path from asynchronously overlapping, causing communication and computation between the end and cloud to become blocked, resulting in an average throughput decrease of 38\% and a 45\% increase in end-to-end latency. These results indicate that both designs play a critical role in supporting system performance, collectively driving comprehensive improvements in performance and scalability of MoE inference.

\section{Conclusion} \label{Conclusion}
In this paper, we propose \textit{EC$^2$MoE}, the first framework to enable adaptive MoE inference via end-cloud pipeline collaboration. This work introduces a novel system architecture that jointly considers expert scheduling, communication efficiency, and hardware heterogeneity. Specifically, we design a hardware-aware lightweight group gate network for efficient and accurate expert routing in end-cloud systems. And we then develop a pipeline optimization mechanism that coordinates inference execution across end and cloud through low-rank compression and route-aware heuristic scheduling. Extensive experiments have shown that \textit{EC$^2$MoE} significantly improves throughput and accuracy while reducing end-to-end latency. At the same time, it also maintains competitive scalability under dynamic workloads and network conditions.

\bibliography{aaai2026}

\end{document}